\documentclass[10pt,a4paper]{article}

\usepackage{color}
\usepackage{textcomp}
\usepackage{stmaryrd}
\usepackage[french]{babel}
\usepackage[latin1]{inputenc}
\usepackage{graphicx,epsfig}
\usepackage{hyperref}
\hypersetup{colorlinks=true, citecolor=black, linkcolor=black, urlcolor=blue}
\usepackage{amssymb}
\usepackage{amsmath}
\usepackage{lastpage}
\usepackage{multicol}
\usepackage{fancyhdr}
\usepackage{url}
\usepackage{makeidx}
\usepackage{caption}
\usepackage{subfig}
\newcommand{\be}{\begin{equation}}
\newcommand{\ee}{\end{equation}}

{\begin{quote}\ttfamily}%
{\end{quote}}

{\begin{quote}\sloppy\itshape}%
{\end{quote}}

\newcommand{\Xa}{\mathbf{X}^a}
\newcommand{\Xb}{\mathbf{X}^b}

\newcommand{\Xt}{\mathbf{X}^t}
\newcommand{\Xx}{\mathbf{X}}
\newcommand{\Yo}{\mathbf{Y}^o}
\newcommand{\Yy}{\mathbf{Y}}

\newcommand{\eb}{\mathbf{\epsilon}^b}

\newcommand{\eo}{\mathbf{\epsilon}^o}

\newcommand{\oB}{\mathbf{B}}

\newcommand{\oH}{\mathbf{H}}

\newcommand{\oK}{\mathbf{K}}

\newcommand{\oR}{\mathbf{R}}

\begin{document}

% \begin{frontmatter}

\title{Reconstruction by data assimilation of the inner temperature field from outer measurements in a thick pipe}

% \author[edfrd]{Jean-Philippe Argaud}
% \ead{jean-philippe.argaud@edf.fr}
% 
% \author[edfrd]{Bertrand Bouriquet}
% \ead{bertrand.bouriquet@edf.fr}
% 
% \author[edfrd]{Aimery Assire}
% 
% \author[edf7n]{St{\'e}phane Gervais}
% 
% \address[edfrd]{
% Electricit\'e de France, R{\&}D
% 1 avenue du G\'en\'eral de Gaulle,
% F-92141 Clamart Cedex - France
% }
% \address[edf7n]{
% Electricit\'e de France, SEPTEN
% 12 avenue Antoine Dutrievoz,
% F-69100 Villeurbanne Cedex - France
% }
% \date{\today}

\author{Jean-Philippe Argaud $^1$ \footnote{jean-philippe.argaud@edf.fr}
   \and Bertrand Bouriquet $^1$ \footnote{bertrand.bouriquet@edf.fr}
   \and Aimery Assire $^1$
   \and St{\'e}phane Gervais $^2$}

\maketitle

\footnotetext[1]{
Electricit\'e de France, R{\&}D
1 avenue du G\'en\'eral de Gaulle,
F-92141 Clamart Cedex - France
}
\footnotetext[2]{
Electricit\'e de France, SEPTEN
12 avenue Antoine Dutrievoz,
F-69100 Villeurbanne Cedex - France
}

\begin{abstract}
The detailed knowledge of the inner skin temperature behavior is very important to evaluate and manage the aging of large pipes in cooling systems. We describe here a method to obtain this information as a function of outer skin temperature measurements, in space and time. This goal is achieved by mixing fine simulations and numerical methods such as impulse response and data assimilation. Demonstration is done on loads representing extreme transient stratification or thermal shocks. From a numerical point of view, the results of the reconstruction are outstanding, with a mean accuracy of the order of less than a half percent of the temperature values of the thermal transient.  
% \end{abstract}

% \begin{keyword}
{\bf keyword:}  

Thermal loading, Heat exchange, Thermal stratification, Thermal shock, Regularization, Impulse response, Inverse problem, Field reconstruction

% \PACS 

% \end{keyword}

% \end{frontmatter}
\end{abstract}

%-------------------------------------------------------------------------------
\section{Introduction \label{sect::intro}}

To improve the status monitoring on main pipes in the cooling system of a nuclear plant, the thermal stress has to be observed through temperature measurements.

\begin{figure}[!ht]
\begin{center}
 \includegraphics[width=0.45\textwidth]{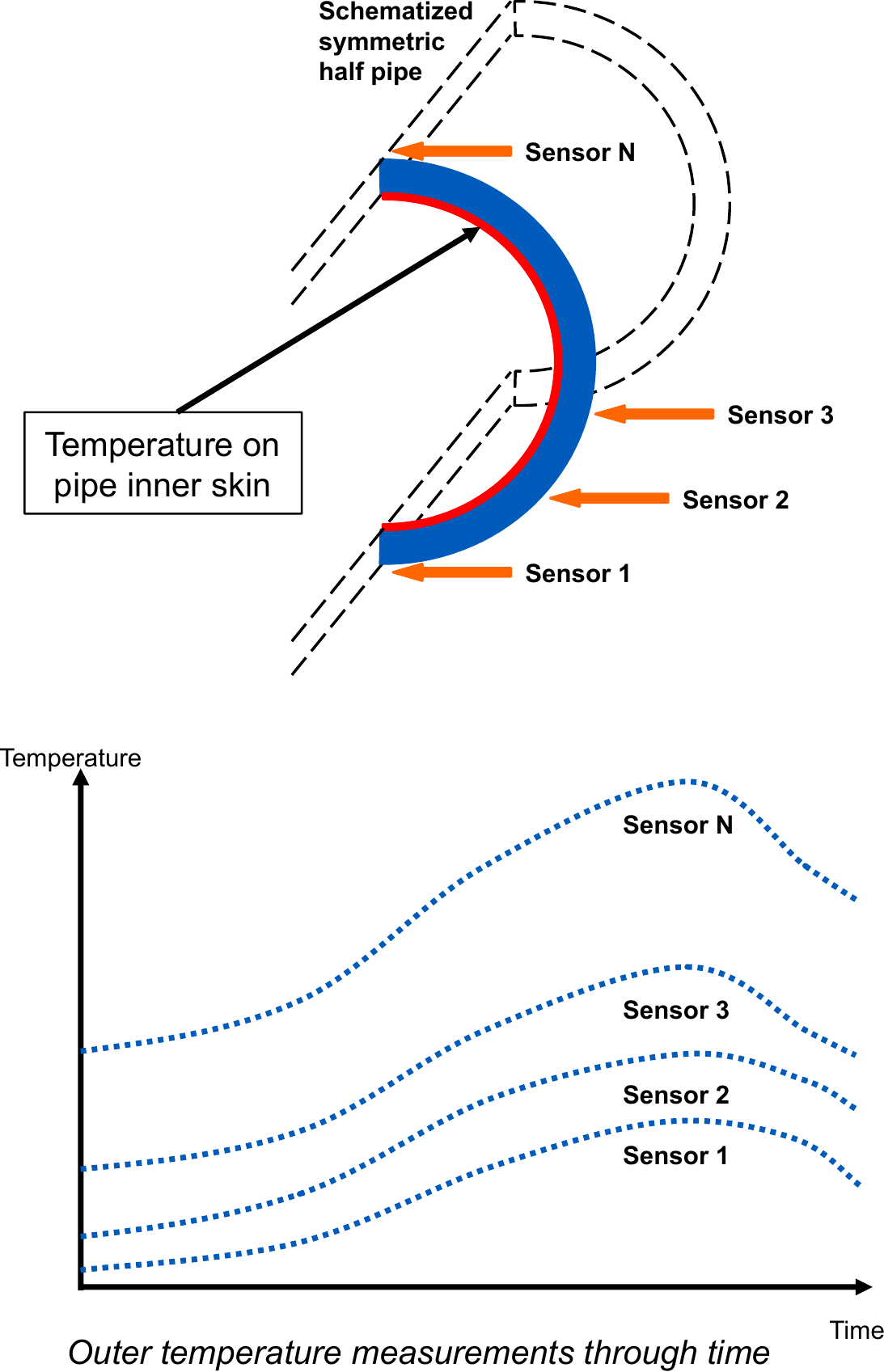}
 \caption{Pipe outer measurements using a ring of devices, on a schematized half pipe. Each device measures the outer temperature through time, and we look for the temperature on the pipe inner skin. \label{fig:pipemeasur}}
\end{center}
\end{figure}

In order to monitor more precisely the aging of the pipe, it is mandatory to get the temperature field on the pipe inner skin. To deal with thermal stratification or thermal shocks in the pipe, originating from inner fluid temperature variations in space and time, new measurement devices have been recently conceived by EDF (see figure \ref{fig:pipemeasur} for a schema of principle, or \cite{Naudin2013} for details). Since no instrument has to be located inside the pipe (that is, on the pipe inner side), both for design reason (least possible hole to deal with the wire) and security reason (least possible element that can be lost in the inner flow), the instrumentation is located outside of the pipe. Moreover, such measurement techniques, with the instrumentation outside of the pipe, it allows to use this system in the current power plants without changing the pipes, because only extra external measurement devices has to be set on each measurement area.

This information can be obtained by using inverse methods, inferring the inner skin temperature from the outer skin temperature measurements (for theoretical background, see for example \cite{Tarantola05}).

The physical conditions are characterized by very large variations of temperature, heterogeneous inner temperature loads, non-linear metallic pipe thermal behavior due to diffusion coefficient depending on temperature, and great thickness of the pipe which leads to signal filtering. Then simple methods are inadequate to solve the problem, and lead to insufficient or aberrant results. As first example of these difficulties, direct inversion of the non-linear simulation of temperature diffusion through the pipe leads to infinity of solutions (or non-physical solutions). Another example, local inversion of temperature, taking into account instrument one by one, independently from the others, is not successful to identify complex vertical stratification or dynamic loads, because of the variable heat diffusion time in the thickness of the pipe.

Then it is necessary to introduce improved inverse methodology, dealing with more realistic physical conditions, and leading to the determination of the internal temperature, in the whole inner skin, as a function of all the known external measurements, both in space and time (thus speaking here of 2D space/time approaches).

We propose here an advanced method that combines data assimilation method and impulse response techniques. Data Assimilation is a robust and efficient method to regularize the physical inversion problem. This technique is widely used in geophysics (see \cite{Bouttier99, Kalnay03, Talagrand97}) and more recently in nuclear engineering (see \cite{Bouriquet2012, Bouriquet2010, Bouriquet2011}) or other industrial domains (see \cite{Argaud09d}). Impulse response is a method, known in numerical and control domains, to reduce the calculation requirements to represent operators in numerical simulations, such as the thermal one we need to use here.

In this paper, we summarize in the first section the data assimilation techniques we use, then we describe in the second section the construction of the required linear operator by impulse response techniques, and we show some significant results in the last section.

To perform the numerical simulation for the thermal problem and the data assimilation methods, we use the \textit{Code\_Aster} thermo-mechanical reference solver and the ADAO module within the SALOME framework (see \cite{AdaoEN}, \cite{Salome}, \cite{SalomeMeca}). \textit{Code\_Aster} solver allows us to get a very good modeling of the non-linear heat diffusion through the metal of the pipe, and is certified for nuclear analysis.

%------------------------------------------------------------------------------
\section{Regularization of the problem by data assimilation \label{sec:regAD}}

We briefly introduce data assimilation key points, to understand their use as applied here. But data assimilation is a wider domain, and these techniques are for example the pillars of current meteorological operational forecast. This is through advanced data assimilation methods that long-term forecasting of the weather has been drastically improved in the last 35 years. Forecasting is based on all the available data, such as ground and satellite measurements, as well as sophisticated numerical models. Some interesting information on these approaches can be found in the following basic already cited references \cite{Bouttier99, Kalnay03, Talagrand97}. 

The ultimate goal of data assimilation methods is to provide a best estimate of the (inaccessible) ``true'' value of a system state, denoted $\Xt$, with the $t$ index standing for ``true''. The basic idea is to put together information coming from an \textit{a priori} state of the system (usually called the ``background'' and denoted $\Xb$), and information coming from measurements (denoted as $\Yo$).  The result of data assimilation is called the analyzed state $\Xa$ (or the ``analysis''), and it is an estimation of the true state $\Xt$ we want to find. Practical and theoretical details on the method can be found for example in \cite{Bouttier99} or \cite{Tarantola05}.

These pieces of information are in relation through models and their numerical simulations. Given the state $\Xx$ of the system, a numerical model of the physical behavior is able to simulate the system in such a way that the output solution $\Yy$ can be compared to observations $\Yo$, when sampled or projected to be in the same space as the observations. The whole process of simulation and sampling, that transform values from the space of the background state to the space of observations, is denoted as the $H$ operator. Its linear approximation is $\oH$, and its reciprocal operator is known as the adjoint of $H$. In the linear case, the adjoint operator is the transpose $\oH^T$ of $\oH$.

Two additional physical information are needed for data assimilation methodology. The first one is the relationships between observation errors at all the measured locations. They are described by the covariance matrix $\oR$ of observation errors $\eo$, considered as random variables, defined by $\eo=\Yo-H(\Xt)$. It is assumed that the errors are unbiased, so that $E[\eo]=0$, where $E$ is the mathematical expectation. The second one is similar, with the relationships between background errors. They are described by the covariance matrix $\oB$ of background errors $\eb$, considered as random variables, defined by $\epsilon_b=\Xb-\Xt$. This represents the \textit{a priori} error, also assumed to be unbiased. There are many ways to obtain this observation and background error matrices. However, they are commonly built from the output of a model with an evaluation of its accuracy, and/or the result of expert knowledge. 

Within this simple framework of state estimation, and in the simplest case where the observation $H$ operator is linear ($H=\oH$), the analysis $\Xa$ is the Best Linear Unbiased Estimator (BLUE), and is given by the following formula: 

\begin{equation}
  \Xa = \Xb + \oK \Big( \Yo - \oH \Xb \Big)
  \label{eqn:BLUEin}
\end{equation}

where the term $\delta\Xa=\oK(\Yo-\oH\Xb)$ is called the analysis increment and $\oK$ is a classical gain matrix (see for example \cite{Bouttier99} for details). This gain matrix can be expressed as a function of covariance matrices of the following form:

\begin{equation}
\oK = \oB\oH^T (\oH\oB\oH^T + \oR )^{-1}
  \label{eqn:gainBLUEbisin}
\end{equation}

It worth noting that, in the case of Gaussian distribution probabilities for the variables $\Xx$ and $\Yy$, solving equation (\ref{eqn:BLUEin}) is equivalent to minimize the following error function $J(\Xx)$, $\Xa$ being its optimal solution:
\begin{equation}\label{eqn:J}
\begin{array}{lcc}
J(\mathbf{x}) &=& (\Xx-\Xb)^T \oB^{-1} (\Xx-\Xb) \medskip \\
              &+& \big(\Yo-\oH\Xx\big)^T \oR^{-1} \big(\Yo-\oH\Xx\big) \\
\end{array}
\end{equation}

This other formulation of the equation (\ref{eqn:BLUEin}) can be used either to solve the problem (for example if it has a large size), or to get some complementary information on the method as presented bellow. 

We can make some enlightening comments concerning this equation (\ref{eqn:J}), and more generally on the data assimilation methodology. If we do extreme assumptions on model and measurements, we notice that these cases are covered by the behavior of the function $J$. First, assuming that the \textit{a priori} state is completely wrong, then the covariance matrix $\oB$ is $\infty$ in quadratic form sense (or equivalently $\oB^{-1}$ is $0$). The minimum of the function $J$ is then given by $\Xa=\oH^{-1}\Yo$ (denoting by $\oH^{-1}$ the pseudo-inverse of $\oH$ in least squares sense). It corresponds directly to information given only by measurements in order to reconstruct the physical field. Secondly, on the opposite side, the assumption that measurements are useless implies that $\oR$ is $\infty$ in quadratic form sense. The minimum of the function $J$ is then evident: $\Xa=\Xb$ and the best estimate of the physical field is then the simulated one from the \textit{a priori} state. Thus, such an approach covers whole range of assumptions we can have with respect to models and measurements. 

In the present case, the state of the system is described by the inner temperature field $\Xx$, which leads through non-linear heat diffusion model to the outer temperature field $\Yo$. This simulation can then be sampled in space and time to be compared to measurement $\Yo$. The background $\Xb$ is an \textit{a priori} guess of the inner temperature field we look for.

There are four required elements to consider to work within the data assimilation formalism:
\begin{itemize}
\item the background \textit{a priori} state $\Xb$,
\item the background error covariance matrix $\oB$,
\item the observation error covariance matrix $\oR$,
\item the linearization of the observation operator, that defines the relationship between the space of the observation and the one of the analysis $\oH$.
\end{itemize}

This background inner temperature $\Xb$ is an important physical hypothesis, and has to be given by expert knowledge. A reasonable assumption is that, after a given delay of stable regime, the outer skin temperature will be equal to the inner skin temperature. As a consequence, as we process a whole time window, the background field $\Xb$ is taken to be the external measurement $\Yo$ with a time delay $\Delta t$.  This delay can be inferred with the typical time transfer. The value is known either through physical experiment, or through numerical impulse response (see next section). It is obvious that this guess $\Xb$  is unacurate, but we know it and then we can set up an error matrix $\oB$  with respect to this knowledge. Morover the situation on the observation errors, given by the matrix $\oR$ in not obvious either. It is worth to recall that the observation error contains not only the measurement errors, but also take into account how well the whole phenomena is represented by the model. Thus this is not, strictly speaking, only the error on the thermocouple measures that has to be taken into account. 

In the present case, we take both matrix $\oB$ and $\oR$ to be diagonal, that in the most neutral hypothesis. This is the best assumption when no other information are known or available. Moreover, as the measurement devices and the inner temperature are homogeneous, the same value of error can be taken over the entire diagonal of both matrices.  Thus, in this case, the main point for optimization is the ratio between the diagonal value of  $\oB$ and the one of $\oR$.  Here, we chose an strong ratio between them, such that $\oR<<\oB$ in quadratic form sense. Fundamentally, this mean that the information coming from observation is far more accurate that the one coming from the background.

The most difficult object to get is the $\oH$ operator, that make the link between the inner temperature states space and the one of observation. With the inversion of this operator, we can describe the variation of the inner temperature as a function of the external one. Its determination is complex, and details are given in the following section \ref{sec:H}.

This approach of our problem under the data assimilation framework can be seen in an equivalent way as a generalized Tychonov regularization, which equations are equivalent (see \cite{Tarantola05} for details). One advantage of the data assimilation framework is to make some explicit links between parametrization choices and physical quantities, which make the setting up easier.

%------------------------------------------------------------------------------
\section{Linear observation operator through impulse response \label{sec:H}}

As required by the equation (\ref{eqn:gainBLUEbisin}), we have to elaborate a linearized expression $\oH$ of the non-linear observation operator $H$. This section recalls the importance of the observation operator and describes the linearization process.

The observation operator allows to model the transformation of a temperature field on the pipe inner skin in an external skin temperature field, that can be measured for example by thermocouples. This operator is non-linear in the present case, mainly because of the temperature dependent heat diffusion behavior of the pipe. Moreover, the temperature evolves during time, so this is an evolution operator. The outer temperatures are sampled in space when measured by thermocouples, and sampled in time by nature of numerical acquisition. So the temperature observations on a time window can be considered together in a unique space/time vector, allowing to process all the measurement devices together, and at once, in the data assimilation inversion. If we sample also the inner temperature, which is natural in numerical discrete simulations, both temperature fields can be seen as 2D space/time vectors. Moreover, the relation between inner and outer temperatures can then be expressed for the whole time window using an adequate $H$ observation operator. Denoting the inner temperatures $\Xx$ and the outer $\Yo$ as above, this leads formally to:

\begin{equation} \label{eqn:yhx}
\Yo = H(\Xx)
\end{equation}

In practice, we use the \textit{Code\_Aster} solver (see \cite{SalomeMeca}) that has non-linear thermal simulation capabilities. The sampling of the inner skin temperature is chosen to be the same as the outer one, that is to say we choose points on the inner skin in front of the outer thermocouple locations. We use the same sampling rate as for the observation (remark that it does not mean that the interior of the pipe wall has to be meshed with this space sampling for the heat equation resolution, it is simply independent). With these choices, the dimension of the $\Xx$ and $\Yo$ discrete representation of the temperature fields are the same, however it is not a requirement of the method.

We will denote respectively $\Yo_{sta}$ and  $\Xx_{sta}$ the initial state of the value $\Yo$ and  $\Xx$ respectively, that will be used  with the property of stationary in time of the $H$ operator. Also we will denote  $\Yo_{delta}$ and  $\Xx_{delta}$ the differences between the stationary state and the current state, in such a way that we can write:

\begin{equation}\label{XYincr}
\begin{array}{lcc}
\Yo &=& \Yo_{sta} +\Yo_{delta} \\
\Xx &=& \Xx_{sta} +\Xx_{delta} 
\end{array}
\end{equation}

At equilibrium between the inner temperature and the outer temperature we got:

\begin{equation} \label{eqn:stat}
\Yo_{sta} = \Xx_{sta} = H(\Xx_{sta})
\end{equation}

This means that, if we are in an equilibrium state, which mean that the temperature is constant with respect to time (but not necessarily with respect to space), the outer temperature is equal to the inner one. Thus we got  $\Yo_{delta} = \Xx_{delta} = 0$. In this case, $H$ become a function without any effect, the inner temperature being strictly equal to the outer one.  In the non stationary case, the code provide temperature and not temperature variation thus we got:

\begin{equation} \label{eqn:lin}
\Yo_{sta} +  \Yo_{delta} = \Yo = H(\Xx) = H(\Xx_{sta} +\Xx_{delta} )
\end{equation}

The knowledge of $\Xx_{delta}$ allows to obtain the value of $\Xx_{sta} +\Xx_{delta} = \Xx$ the inner skin temperature. The key point is then in the determination through inversion of $\Xx_{delta}$ from a  $\Yo_{delta}$. In order to perform this inversion through data assimilation, it is mandatory to build a linearization, noted $\oH$, of the $H$ operator.  Though this linear approximation, we can write:

\begin{equation} \label{eqn:statlin}
 H(\Xx_{sta} +\Xx_{delta})  \approx H(\Xx_{sta}) +\oH \Xx_{delta}
\end{equation}

Using the equation (\ref{eqn:lin}( and the stability at equilibrium expressed by the condition (\ref{eqn:stat}), we got:

\begin{equation} \label{eqn:delta}
\Yo_{delta} = \oH \Xx_{delta} 
\end{equation}

We can write formally that ``$ \Xx_{delta} =\oH^{-1} \Yo_{delta}$''. Such a formulation shows clearly that the knowledge of an increment of temperature $\Yo_{delta}$ around a known $\Yo_{sta}$ outer state allows to calculate the variation $\Xx_{delta}$ of temperature of the pipe inner skin around the corresponding $\Xx_{sta}$ inner known state.  To make the notation smoother, we will then remove the $delta$ index in the remaining part of the paper, as only variation are taken in into account by $H$ operator.

Because the formal inversion stated above is ill-posed, it is required to regularize the problem by using data assimilation methodology, to obtain a well-behaved $\Xa$ solution. In particular, using equation (\ref{eqn:gainBLUEbisin}) requires this linear operator $\oH$.

The simplest idea to determine $\oH$ is to build it by finite differences of $H$ with respect to the initial state $\Xx_{sta}$. In physical terms, it requires the incremental response for each temperature variation at each time on each measurement device.

With such a procedure, we do obtain the linear operator $\oH$ as a matrix that gives the change of temperature on outer skin, for each time on each measurement devices, for a given change of inner skin temperature. However, in most cases, this method is computationally exceedingly costly. For example, assuming commonly that we got $10$ devices observed on a time window of $1000$ steps, it will require $10~000$ unitary non-linear calculations to build the whole linear operator. Assuming that a unitary calculation last for $1$  minute, the total computational time requirement for the linear $\oH$ matrix would last for about one full week, which is far too much. Thus, a more efficient solution has to be found, and in this purpose, we use the impulse response of the pipe with respect to a sudden thermal inner variation. We shortly describe this classical method, related to Green functions and fundamental solutions of differential equations (see \cite{Roach1982,Stakgold2011} for detailed information).

Let's use a thermal transient, described by the temperature change  $\Xx$ as a function of space and time. As shown on figure \ref{fig:pipemeasur}, the measurement devices are located on a ring around the pipe, irregularly spaced, so the angle location of any point on the ring is sufficient to describe either an inner or an outer temperature location. We look for the inner temperature also on an internal ring. It is discretized in space using $p$ angles numbered by $[1...p]$, and in time over a time window of $n$ time steps named $[t_1...t_n]$. We will use the Kronecker functions (in space or time) such that $\delta_k^i =0, \forall i\neq k$ and $\delta_k^k =1$. Then, by definition, we can decompose the function $X(\theta,t)$  in the discertized space in the following way:

\begin{equation} \label{eqn:ti}
\forall m \in [1...p], j \in [1...n], X(m,j) = \sum_{l=1,p} \sum_{i=1,n} \Xx_{li}  \delta_l^m \delta_i^j
\end{equation}

The scalar value $X_{li}$, that multiplies the Kronecker functions product, represents the temperature at time $t=t_i$ in the chosen location $\theta=l$. It is obvious that all the values of $\Xx$ are exactly represented for all the time $[t_1...t_n]$ on location $[1..p]$ from the function described in equation (\ref{eqn:ti}). By definition of Green functions, which are the impulse responses of the $H$ operator to Kronecker functions, the response of the function $X(\theta,t)$ through the $H$ operator is a convolution, leading to the discrete expression:

\begin{equation} \label{eqn:tid}
\forall m \in [1...p], j \in [1...n], H(X) (m,j) = \sum_{l=1,p} \sum_{i=1,n} \Xx_{li}  H(\delta_m^l \delta_j^i)
\end{equation}

This is the classical discrete transformation of $H(\Xx)$ though  the elementary transformation of the pulse in space and time. Those value are the response to impulse at location $m$ and at time $j$ to a pulse in inner skin. 

In other words, an impulse response is a transfer function that put in relation a pulse done at position $l$ at time $i$ with what is seen by the measurement device at a location $m$ et time $j$. This impulse response is given by the coefficients $H(\delta_m^l \delta_j^i)$, denoted as $C^{li}_{mj}$ to insist on its constant nature with respect to $\Xx$. If the value of the transformation of the state $\Xx$ through $\oH$ at location $(m,j)$ is denoted $\Yo_{mj}$, by analogy with the continuous formulation, we got:

\begin{equation} \label{eqn:tid2}
\Yo_{mj} = \sum_{l=1,p} \sum_{i=1,n} \Xx_{li}  C^{li}_{mj} 
\end{equation}

The practical key point is the calculation of the elementary response functions $C^{li}_{mj}=H(\delta_m^l \delta_j^i)$. In the discrete equation (\ref{eqn:tid2}), these functions give access to the behavior of the system to an inner thermal loading at point $l$ and time $i$ seen by a measurement device at point $h$ and time $k$. Due to the time translation properties of the operator $H$, it is only the difference $t_i-t_k$ (where $t_i>t_k$) between the time steps $i$ and $k$ that matters, not the absolute time values of $t_i$ or $t_k$.  Thus we can write:

\begin{equation} \label{eqn:tid3}
\Yo_{mj} = \sum_{l=1,p} \sum_{i=1,n} \Xx_{li}  C^{li}_{m(j-i)} 
\end{equation}

It reduces the number of required constants by using the same one for each identical difference between time steps $i$ and $j$. It has been verified that the impulse responses can be superposed (the response to a double impulse is the sum of the response to each impulse done separately), are symmetric (the responses at symmetric locations around an impulse location are the same) and are maximal on the closest measurement device corresponding to the impulse but with a delay depending on the thickness of the pipe. These properties prove that we can effectively split the internal solicitation, as it is done in the equation (\ref{eqn:tid}) for the impulse response.

So the constants $H(\delta_m^l \delta_j^i)$ has only to be calculated for a Kronecker impulse at each discrete angle on the inner ring and at the first time step, obtaining the outer temperature fields at all the discrete angles and at all the time steps. These constants are the elementary response matrix for Kronecker impulses. These elementary response calculations can be gathered in order to build the linear relation expressed in equation (\ref{eqn:tid3}), in such a way that the inner skin temperatures $\Xx_{li}, \forall l, i$, considered as a vector, can be algebraically multiplied by a matrix. It defines the $\oH$ operator, build by stacking shifted elementary response matrix. On overall, the number of elementary impulse response calculations is reduced to only one calculation per angular position of measurement devices, where we have information. To compare to the simplest finite difference idea for $\oH$ indicated above, where we need $10~000$ unitary non-linear calculations, here we only need $10$ unitary non-linear calculations to build the whole $\oH$ linear operator, independently of the number of time steps.

\begin{figure}[!ht]
\begin{center}
 \includegraphics[width=0.95\textwidth]{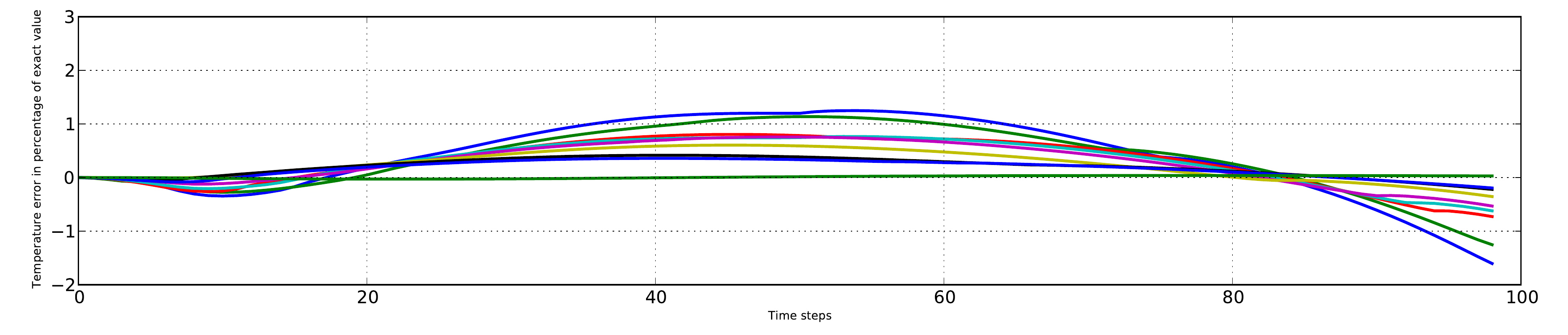}
 \caption{Differences between the outer skin temperatures obtained either through a non-linear calculation $H$, or through the approximation obtained thought the linear matrix $\oH$ build by impulse responses. \label{fig:linqual}}
\end{center}
\end{figure}

To illustrate the quality of the linearized operator $\oH$ with respect to the non-linear one $H$, we check the already stated relation (\ref{eqn:tid3}) by comparing both the linear and non-linear calculations, on a particular case that will also be used in the next result section \ref{sec:results} and for the figure \ref{fig:results}. The differences, on measurement points along time, are presented in the figure \ref{fig:linqual}, in percentage of the non-linear reference calculation. It shows an excellent agreement between the approximation and the exact calculation, with less than $1.5\%$ of maximum error over the whole time window, and far less on mean. Then, using the impulse response methodology, an excellent linearized observation operator can then be build at a cheap computational price.

We now have all the required ingredients to set up the standard data assimilation regularization methodology using the ADAO module within the SALOME framework, building the linear operator by using the \textit{Code\_Aster} solver on adequate finite elements meshing for the pipe wall.

%------------------------------------------------------------------------------
\section{Results \label{sec:results}}

For the present demonstration, we assume that there are $9$ external measurement devices, irregularly located on the half symmetric ring at various incremental angles. This choice of number $9$ is perfectly arbitrary, and do not reflect design choices done for the effective devices. Moreover, the inverse method itself can take into accounts more or less measurements devices. We choose to use a twin experiments framework to assess the quality of the reconstruction: making an hypothesis $\Xt$ on a given inner skin temperature field, we simulate the outer field and sample it to get pseudo-observations $\Yo$. Then we apply the data assimilation methodology using these pseudo-observations to reconstruct an inner temperature field $\Xa$. Finally, this result can be compared to the initial hypothesis $\Xt$ (which is impossible in case of real measurements, as the true value is unknown, and only available in twin experiments).

It is important to note that the ``true'' reference value $\Xt$ is chosen here to be extreme, in the sense that the sudden changes in time of slope of the temperature curves are not physically feasible. However, here, we are only focusing on the efficiency of the method, so this is not an issue to propose complex or extreme non-physical state. If such a state can be handled, then simplest or smoothest states are easiest to deal with.

\begin{figure}[!ht]
\begin{center}
 \includegraphics[width=0.95\textwidth]{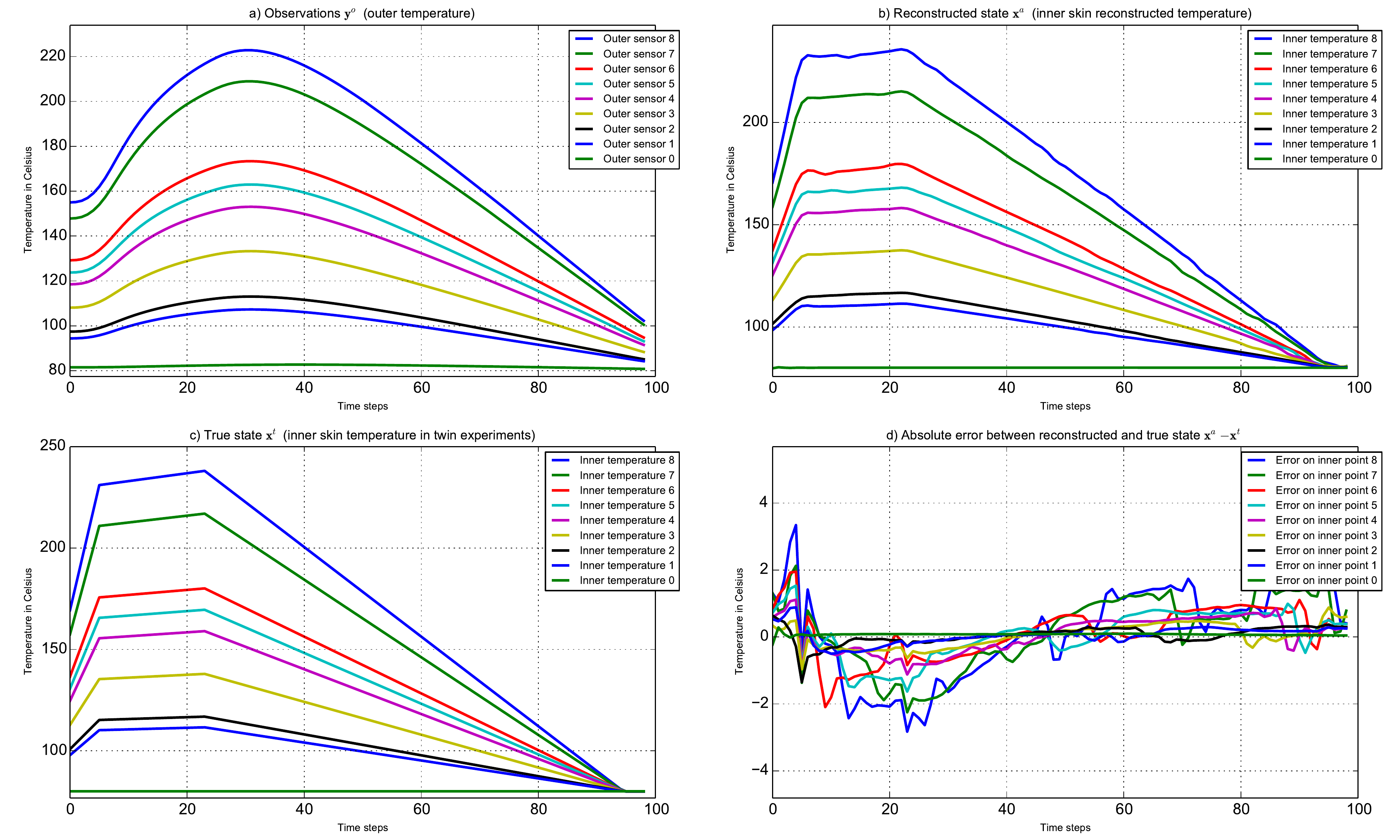}
 \caption{Stratified case: reconstruction of the inner skin temperature $\Xa$ from observations $\Yo$, and comparison to true value $\Xt$ (amplified scale for the error) in the case of a stratified transient.\label{fig:results}}
\end{center}
\end{figure}

On the figure \ref{fig:results} are presented the $\Yo$ data seen by the measurement devices on top left panel $a$, the result $\Xa$ of the reconstruction on top right panel $b$, the inner temperature field $\Xt$ used for the transient in twin experiments on bottom left panel $c$,  obtained by processing of the measurements on the panel $a$. Finally, on bottom right panel $d$, are presented the differences $\Xa-\Xt$ between results of the reconstruction (panel $b$) and the reference value (panel $c$).

Comparing the results $\Xt$ of the panel $b$ and $\Xa$ on the panel $c$, it is clear that the results are outstanding. This can be more clearly seen on the panel $d$, were the differences are drawn (using a magnification of a factor of about 40 with respect to the three other sub-figures). On the mean, the difference is of less than $0.4\%$ of the temperature value, which is excellent. Moreover, the maximum of absolute error is only about $3^{\circ}$C (at a time step which is not meaningful because of temperature slope change, see below), to be compared to temperature values ranging from $80^{\circ}$C to about $240^{\circ}$C. Finally, the order of the thermal transient curve panel $c$ in figure \ref{fig:results} is then completely coherent with the original one.

We notice that the maximum of error is located on the slope change location that, in the chosen transient, is sharp as stated above. This small effect has been studied in details in several other cases and can be understand theoretically. Results in physical cases are then far better, even taking into account measurement noise. 

\begin{figure}[!ht]
\begin{center}
 \includegraphics[width=0.95\textwidth]{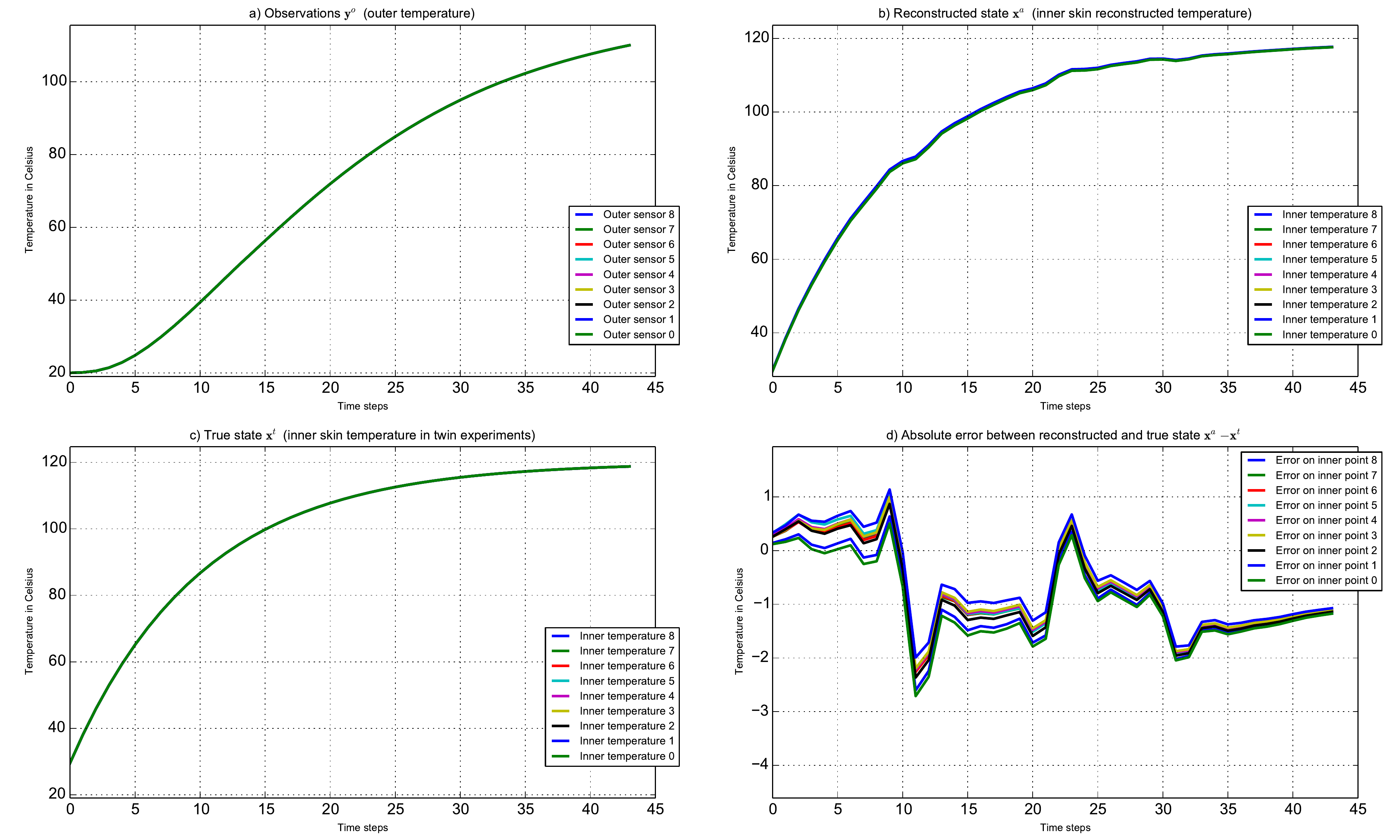}
 \caption{Shock case: reconstruction of the inner skin temperature $\Xa$ from observations $\Yo$, and comparison to true value $\Xt$ (amplified scale for the error) in the case of a thermal shock. \label{fig:choc}}
\end{center}
\end{figure}

In order to go further in the method test, we use a more stringent situation as in the previous case. One classical case is a thermal shock, when there is a sudden rising of the temperature all over the pipe. Then the temperature goes quickly from a cold state to a hot one, at all measurement points. The shape of the transient is presented on the panel $c$ of the figure \ref{fig:choc}.

As it can be seen on figure \ref{fig:choc}, the chosen shock is rather strong. It consists in a fast temperature increase of $100^{\circ}$C in a short time, which is more stiff than what is done in the case of figure \ref{fig:results}. In this case, the measurements on outer skin presented on the panel $a$ are mainly a delay on the shock, with only a change in shape with respect to the true state, which make the curve smoother dur to thermal diffusion property of the pipe. Using the reconstruction method, we notice on the panels $c$ and $d$ that the initial shape is well reconstructed.  

The error is at most $2.6^{\circ}$C in such an extreme case of reconstruction, as it can be seen on the panel $d$ of figure \ref{fig:choc}. All over the considered time window, the mean error is about $0.9\%$.

It can be noticed on the panel $d$ of figure \ref{fig:choc} that there are some discrepancies, of at most a few tenths of percents, between the various device locations where the reconstruction is done. In theory, as the initial shock is homogeneous in temperature, all the curve should be identical as shown on the panels $c$ or $a$. Those differences are linked to the angular irregular locations of the instruments. This is in fact related to the delay of the thermal exchange between the various devices, as outer temperature is the sum of thermal diffusion from all the inner sources around the pipe, coming with variable time delays. As a consequence, the central instrument, that is perfectly symmetric in angle difference with respect to the other ones, gives the best result. The gathered information is well synchronized in time with respect to the two top and bottom sides. Such effect disappear if all the measurement devices are evenly located, so no differential delay exists between the contribution of symmetric sources. However, this effect can be considered as negligible.

To model a thermal shock, we did not chose a step function to keep some physical meaning in the transient. Such step effect cannot be seen in operation. However, numerically, it can be done and we evaluate the discrepancy of the method. The step case lead only to a maximal local error of $4.5\%$, in the same computation conditions as the ones used in the case presented in figures \ref{fig:results} or \ref{fig:choc}, pointing out the change in slope. The rising speed chosen was equivalent to $60^{\circ}\text{C}.s^{-1}$, as the time is discrete. 

\begin{figure}[!ht]
\begin{center}
 \includegraphics[width=0.95\textwidth]{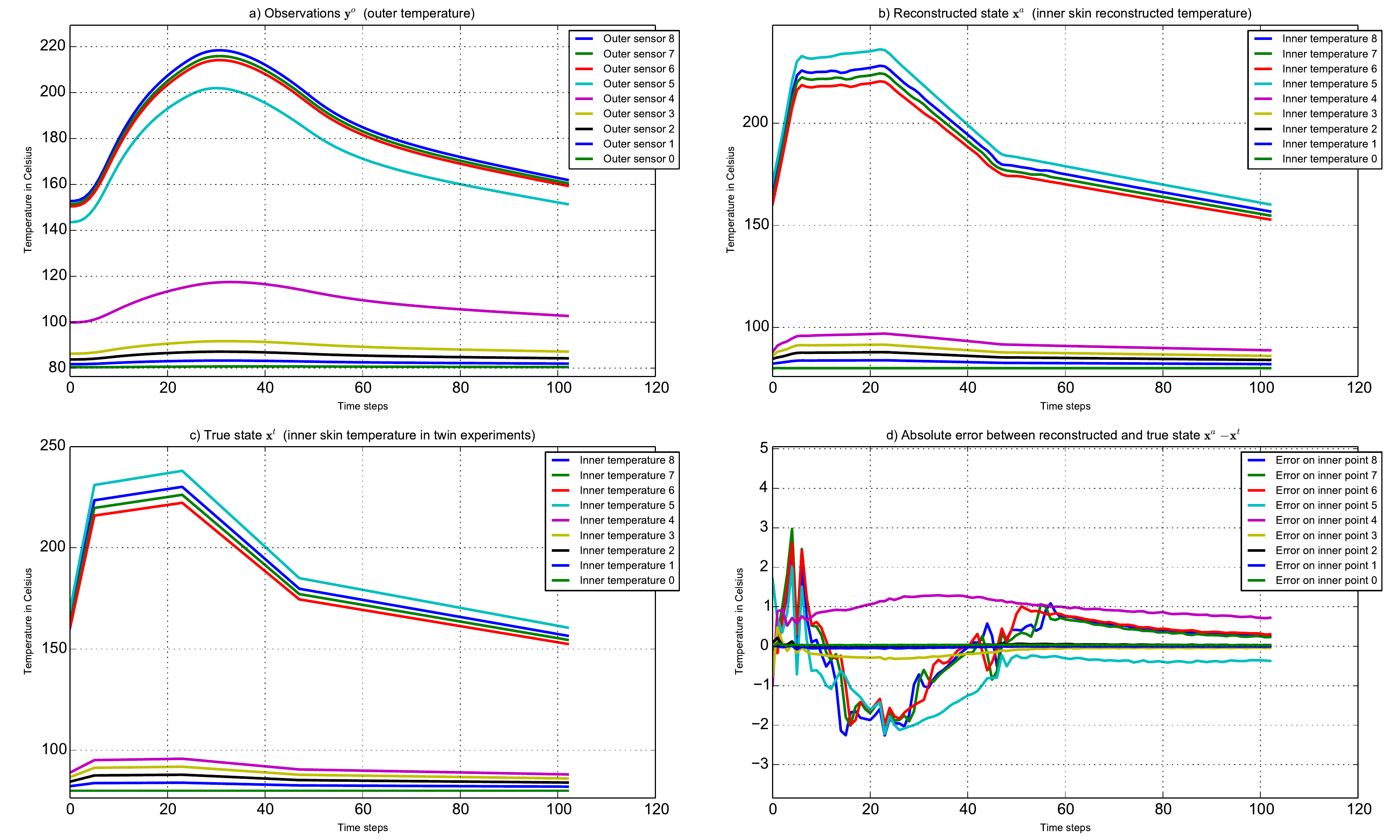}
 \caption{Inversion case: reconstruction of the inner skin temperature $\Xa$ from observations $\Yo$, and comparison to true value $\Xt$ (amplified scale for the error). \label{fig:inversion}}
\end{center}
\end{figure}

In the shock case shown in figure \ref{fig:choc}, it may appear that taking the whole information from all the measurements is not useful, as all the thermal layer are the same at each time.  It can be tempting to use a one dimensional inversion model (1D inversion: for one given device location, inversion of time varying measurements), even if it is clear there are connections between the different source locations through thermal diffusion in the pipe. But this is this non-instantaneous diffusion that makes impossible to use a one dimensional inversion model. To understand better this effect of thermal diffusion inside the pipe, we study here a peculiar case.  The observation associated are shown on the panel $a$ of figure \ref{fig:inversion}.

The first guess, with respect to the outer measurement pattern shown on the panel $a$ of the figure \ref{fig:inversion}, is that there is a similar continuous stratification of the inner skin temperature from the bottom to the top. This answer would be the one given by a 1D reconstruction of the inner temperature. However this answer is wrong, as demonstrated by the true state of the inner temperature that is shown on the panel $c$ of the figure \ref{fig:inversion} or the reconstruction on the panel $b$. Actually, there is not a continuous increasing of the temperature from the bottom to the top, but, in fact, there is an inversion of the temperature around the middle of the pipe. This is the cross exchange of heat in the pipe that is finally contradictory with the illusion of a continuous increasing of the temperature on the inner skin, by similarity with outer temperatures on the panel $a$.  

\begin{figure}[!ht]
\begin{center}
 \includegraphics[width=0.95\textwidth]{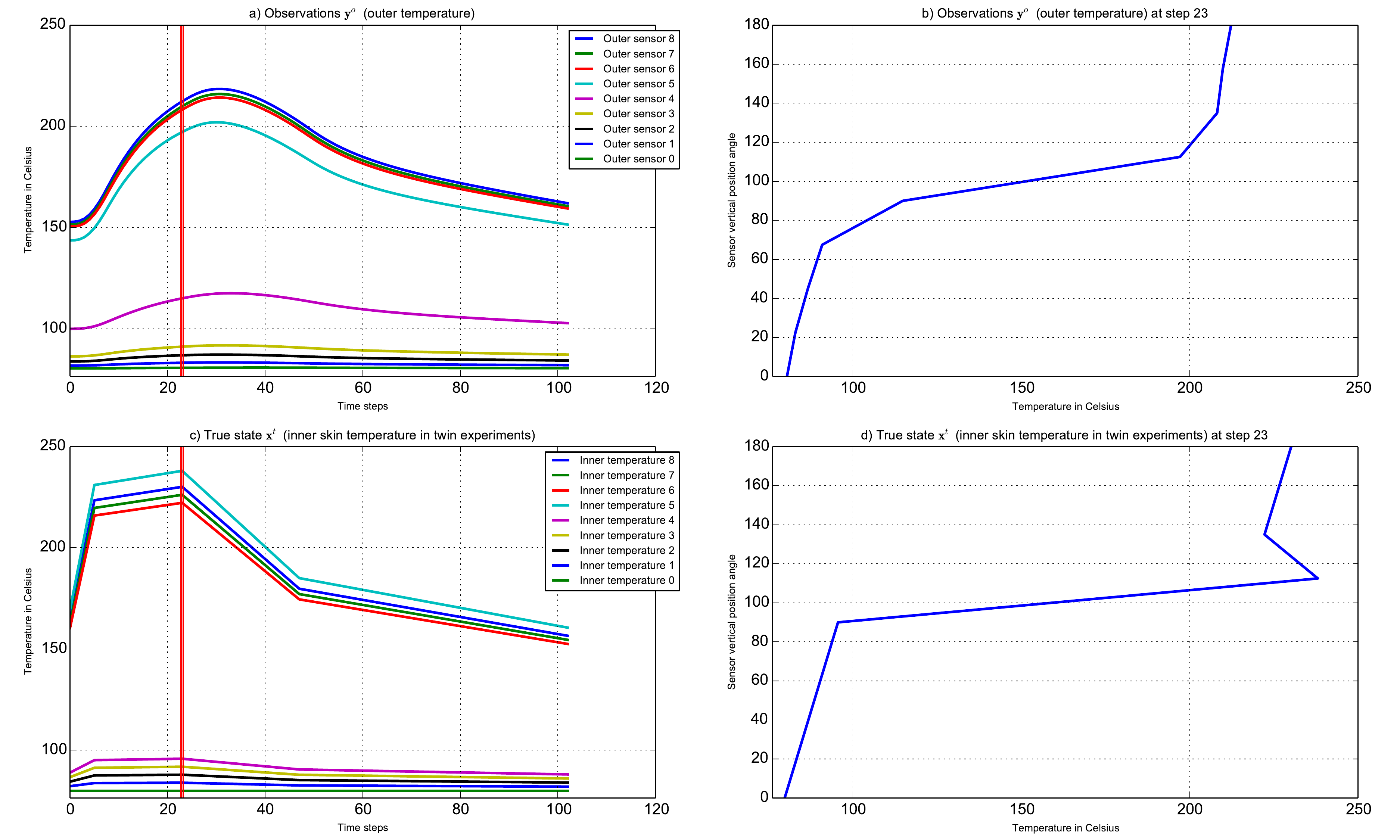}
 \caption{Evolution of the temperature as a function of the angle for the outside measurements and the true inner values (or the reconstructed values, which are similar to the true ones). \label{fig:inversioncut}}
\end{center}
\end{figure}

This can be seen in the figure \ref{fig:inversioncut}, that show the evolution of temperature as a function of the angle, for one arbitrary time step (marked by a red line) corresponding for the outside measurement and the true value. On the figure \ref{fig:inversioncut}, both outer and inner temperatures are represented in upper and lower part respectively.  In this figure the left side represent the time evolution on the various location, and the right side the evolution of temperature as a function of the angle.

For the considered arbitrary time step in figure \ref{fig:inversioncut},  it is clear that, with respect to the outer data, there is a regular increasing of the temperature from bottom to top, as it can be seen on the panel $b$.  However the true state present a clear inverted shape around the middle of the pipe, shown on the panel $d$. And this peculiar shape of the inner temperature is perfectly reconstructed by the proposed methodology, as shown on figure \ref{fig:inversion}.  This excellent quality of reconstruction is available all over the considered time window, as shown on the panel $d$ of figure \ref{fig:inversion}. The error is locally at most of $3^{\circ}$C, and on the mean of $0.3\%$ all over the time window. Those values are very similar to the one obtained in the case of the figure \ref{fig:results} that was the basis used to develop the case of  figure \ref{fig:inversioncut}. This means that  this peculiar behavior has no influence on the reconstruction quality.

This last example demonstrate all the efficiency of the 2D inversion method, and its tremendous gain respect to a 1D inversion method, that is unable to consider correctly stratification or even more complex situations, with heat exchange effects through the structure. 

%------------------------------------------------------------------------------
\section{Conclusion \label{sect::conclusion}}

In this article, we present an advanced compound method of impulse response and data assimilation regularization to obtain the inner temperature, in a thick pipe, from outer temperature measurements, on a time window.

The results of the proposed method are outstanding, far beyond expected ones. The first point is the overall excellent accuracy, that is on the mean of less than $0.4\%$, with a maximal error of about $3^{\circ}$C to be compared to values on the order of $80^{\circ}$C to $240^{\circ}$C even for extreme thermal cases. The second point is that the method deals with the full space/time 2D set of measures, that is mandatory in the operational case, due to the slow heat diffusion that leads to spread valuable information between all the measurement devices. In itself, this method is then a remarkable solution to the problem of inner temperature reconstruction from outside measurements.

The next step is the operational implementation of the method in a complete chain of measurements that will allow a fine monitoring of the state of the pipes.

%------------------------------------------------------------------------------
\bibliographystyle{elsarticle-harv}
\bibliography{RTG}

\end{document}